\documentstyle[12pt]{article}
\catcode`\@=11
\long\def\@makefntext#1{
\protect\noindent \hbox to 3.2pt {\hskip-.9pt
$^{{\ninerm\@thefnmark}}$\hfil}#1\hfill}		

\def\@makefnmark{\hbox to 0pt{$^{\@thefnmark}$\hss}}  
\def\ps@myheadings{\let\@mkboth\@gobbletwo
\def\@oddhead{\hbox{}
\rightmark\hfil\ninerm\thepage}
\def\@oddfoot{}\def\@evenhead{\ninerm\thepage\hfil
\leftmark\hbox{}}\def\@evenfoot{}
\def\sectionmark##1{}\def\subsectionmark##1{}}

\renewcommand{\thefootnote}{\fnsymbol{footnote}}

\newcounter{sectionc}\newcounter{subsectionc}\newcounter{subsubsectionc}
\renewcommand{\section}[1] {\vspace*{0.6cm}\addtocounter{sectionc}{1}
\setcounter{subsectionc}{0}\setcounter{subsubsectionc}{0}\noindent
	{\normalsize\bf\thesectionc. #1}\par\vspace*{0.4cm}}
\renewcommand{\subsection}[1] {\vspace*{0.6cm}\addtocounter{subsectionc}{1}
	\setcounter{subsubsectionc}{0}\noindent
	{\normalsize\it\thesectionc.\thesubsectionc. #1}\par\vspace*{0.4cm}}
\renewcommand{\subsubsection}[1]
{\vspace*{0.6cm}\addtocounter{subsubsectionc}{1}
	\noindent {\normalsize\rm\thesectionc.\thesubsectionc.\thesubsubsectionc.
	#1}\par\vspace*{0.4cm}}

\def\abstracts#1{{
\centering{\begin{minipage}{12.2truecm}\vspace*{.1cm}
        \footnotesize\baselineskip=12pt\noindent
	\parindent=0pt #1
	\end{minipage}}\par}}

\renewenvironment{thebibliography}[1]
	{\begin{list}{\arabic{enumi}.}
	{\usecounter{enumi}\setlength{\parsep}{0pt}
\setlength{\leftmargin 1.25cm}{\rightmargin 0pt}
	 \setlength{\itemsep}{0pt} \settowidth
	{\labelwidth}{#1.}\sloppy}}{\end{list}}

\def\@cite#1#2{\unskip\nobreak\relax
    \def\@tempa{$\m@th^{\hbox{\the\scriptfont0 #1}}$}%
    \futurelet\@tempc\@citexx}
\def\@citexx{\ifx.\@tempc\let\@tempd=\@citepunct\else
    \ifx,\@tempc\let\@tempd=\@citepunct\else
    \let\@tempd=\@tempa\fi\fi\@tempd}
\def\@citepunct{\@tempc\edef\@sf{\spacefactor=\the\spacefactor\relax}\@tempa
    \@sf\@gobble}
     
\def\citenum#1{{\def\@cite##1##2{##1}\cite{#1}}}
\def\citea#1{\@cite{#1}{}}
     
\newcount\@tempcntc
\def\@citex[#1]#2{\if@filesw\immediate\write\@auxout{\string\citation{#2}}\fi
  \@tempcnta\z@\@tempcntb\m@ne\def\@citea{}\@cite{\@for\@citeb:=#2\do
    {\@ifundefined
       {b@\@citeb}{\@citeo\@tempcntb\m@ne\@citea\def\@citea{,}{\bf ?}\@warning
       {Citation `\@citeb' on page \thepage \space undefined}}%
    {\setbox\z@\hbox{\global\@tempcntc0\csname b@\@citeb\endcsname\relax}%
     \ifnum\@tempcntc=\z@ \@citeo\@tempcntb\m@ne
       \@citea\def\@citea{,}\hbox{\csname b@\@citeb\endcsname}%
     \else
      \advance\@tempcntb\@ne
      \ifnum\@tempcntb=\@tempcntc
      \else\advance\@tempcntb\m@ne\@citeo
      \@tempcnta\@tempcntc\@tempcntb\@tempcntc\fi\fi}}\@citeo}{#1}}
\def\@citeo{\ifnum\@tempcnta>\@tempcntb\else\@citea\def\@citea{,}%
  \ifnum\@tempcnta=\@tempcntb\the\@tempcnta\else
   {\advance\@tempcnta\@ne\ifnum\@tempcnta=\@tempcntb \else \def\@citea{--}\fi
    \advance\@tempcnta\m@ne\the\@tempcnta\@citea\the\@tempcntb}\fi\fi}

 1
 1
 1

\font\ninerm=cmr9

\def\half{{\textstyle{1\over 2}}}

\begin{document}
\begin{flushright}
BOW-PH--109 \\
hep-ph/9710437 \\
\end{flushright}
\vspace{0.8cm}

\centerline{\normalsize\bf Fermion Vacuum Effects on Soliton 
Stability\footnote{To appear in the proceedings of
{\it Solitons: Properties, Dynamics, Interactions and Applications}, 
Kingston, Ontario, Canada, July, 1997.}
}

\vspace*{0.6cm}
\centerline{\footnotesize STEPHEN G. NACULICH\footnote{naculich@bowdoin.edu}}
\vspace*{0.3cm}
\baselineskip=13pt
\centerline{\footnotesize\it Department of Physics}
\baselineskip=12pt
\centerline{\footnotesize\it Bowdoin College}
\centerline{\footnotesize\it Brunswick, ME  04011, U.S.A.}
\baselineskip=13pt

\vspace*{0.9cm}
\abstracts{
Just as fermion zero modes can alter the degeneracy 
and quantum numbers of a soliton,
fermion energies can affect the form and 
stability of a nontopological soliton.
We discuss the kink in a two-dimensional linear sigma model,
and show that, when coupled to fermions,
the kink is no longer an extremum of the energy functional.
The kink in this model possesses many similarities to the 
electroweak string in the Weinberg-Salam model.  }

\normalsize\baselineskip=15pt
\setcounter{footnote}{0}
\renewcommand{\thefootnote}{\alph{footnote}}

\section{Nontopological Solitons}
\setcounter{equation}{0}
\renewcommand{\theequation}{1.\arabic{equation}}

A nontopological soliton 
is a nontrivial static configuration of bosonic fields.  
If it is stable, its stability is due to 
energetic rather than topological reasons, 
{\it i.e.}, it is a local minimum of the bosonic energy functional 
\begin{equation}
E_{\rm boson} [\Phi]  = ~-~\int {\rm d}^n x~ {\cal L}_{\rm boson} (\Phi)
\end{equation}
where $\Phi$ schematically represents all the bosonic 
(scalar and vector) fields. 

If fermion fields are coupled to the bosonic fields,
one may solve the Dirac equation 
in the static soliton background to obtain a spectrum
$\epsilon_\lambda [\Phi]$ of positive and negative eigenenergies.
One may then consider excited states of the nontopological soliton, 
which have some of the positive energy (or ``valence'') modes filled. 
These excited states are found by minimizing 
\begin{equation}
E[\Phi] = E_{\rm boson} [\Phi] 
+ \sum_{\epsilon_\lambda > 0} n_\lambda \epsilon_\lambda [\Phi] 
\end{equation}
where $n_\lambda$ is the occupation number of the positive energy modes.
The presence of valence fermions modifies the soliton background $\Phi$.

To take into account the effect 
of occupied valence states on the soliton
while ignoring that of the fermion vacuum energy
({\it i.e.}, the energy of the ``Dirac sea''),
is not consistent, however,  in an $\hbar$-expansion.
A more consistent approach is to minimize,
usually in some approximation, the quantity
\begin{equation}
E[\Phi]= E_{\rm boson} [\Phi] + E_{\rm vacuum} [\Phi]
+ \sum_{\epsilon_\lambda > 0} n_\lambda \epsilon_\lambda [\Phi] 
+ \sum_{\epsilon_\lambda < 0} m_\lambda \left|  \epsilon_\lambda [\Phi] \right|
\label{effectiveenergyxyz}
\end{equation}
where we have also included the possible presence of ``holes,''
or vacancies in the ``Dirac sea'';
$m_\lambda$ is the hole occupation number.

While including fermion loop effects,
we ignore boson loop effects throughout, 
treating bosonic fields as classical.
The legitimacy of this approximation
depends on the particular application.
In any case, this approach can be systemically justified 
through a large $N$ approximation 
($N$ is the number of fermion flavors), 
in which bosonic loops are subleading in $1/N$.

In my talk, I discussed the effects of fermions 
on three types of nontopological solitons:
Higgs bags around heavy fermions,
kinks in a two-dimensional linear $\sigma$-model, and 
electroweak strings in the Weinberg-Salam model.
Here, due to space limitations, 
I will focus on the middle topic,
referring the reader to refs.~[\citenum{BaggerNaculich91,BaggerNaculich92}]
for a discussion of the first topic,
and to refs.~[\citenum{Naculich95,KonoNaculich96,LiuVachaspati96}]
for discussions of the last.

\section{Kinks in the linear sigma model}
\setcounter{equation}{0}
\renewcommand{\theequation}{2.\arabic{equation}}

The two-dimensional linear $\sigma$-model
serves as a toy model to illustrate
some of the features of electroweak strings in the
Weinberg-Salam model.
The electroweak string is a nontopological soliton which,
for a certain range of the Higgs mass and $\sin^2 \theta_W$,
is a saddle point of the energy functional, and therefore is unstable.
The kink configuration in the 2d $\sigma$-model is
also a saddle point.
Both the kink and the electroweak string have fermion zero modes, 
which make them degenerate.
And in both cases, 
perturbations in the field configuration shift the zero mode, 
lifting the degeneracy 
and crucially affecting the stability of the soliton.

The linear $\sigma$-model Lagrangian
\begin{equation}
{\cal L}_{\rm boson}
         = {\half} \left(\partial_\mu  \sigma  \right)^2
         + {\half} \left(\partial_\mu  \tau     \right)^2
	 - {\textstyle{1\over 4}} \lambda  
            \left( \sigma^2 + \tau^2 - v^2 \right)^2
\end{equation}
has a Mexican-hat-shaped potential. 
Consider the set of kink-like configurations
\begin{eqnarray}
&&\tau (x) 	= \tau = {\rm const},
\qquad {\rm where}  -v \le \tau  \le v \nonumber\\
&&\sigma (x) 	= \sqrt{v^2 - \tau^2} ~ \tanh 
\left( \sqrt{1 - {\tau^2 \over v^2}} {x\over x_0} \right)
\label{kinkconfigurationsxyz}
\end{eqnarray}
where $x_0 = \sqrt{2/\lambda v^2}$.
The energy of these configurations is given by 
\begin{equation}
E_{\rm boson}(\tau) = {\textstyle {2\over 3}} \sqrt{2 \lambda} v^3
 \left( 1 - {\tau^2 \over v^2} \right)^{3/2}. 
\label{kinkenergyxyz}
\end{equation}
The configuration with $\tau = 0$ is the usual kink,
which goes over the crown of the Mexican hat;
it is a saddle point of the energy, and maximizes eq.~(\ref{kinkenergyxyz}).
The configurations with $\tau = \pm v$ are vacuum configurations,
and the family (\ref{kinkconfigurationsxyz}) interpolates
between the kink and the vacuum.

We now couple fermions to this model by adding
\begin{equation}
{\cal L}_{\rm fermion} 
= \bar{\psi} \left( i {\rlap{/}\partial} -g \sigma +ig\tau\gamma_5 \right) \psi
\end{equation}
to the Lagrangian.
The Dirac equation has a set of discrete eigenvalues 
\begin{equation}
\epsilon_n = {1\over x_0} \sqrt{ y^2 - (\sin \alpha)^2 (y-n)^2 },~\qquad\qquad
0 \le n < y
\end{equation}
where $y=g\sqrt{2/\lambda}$ and $\cos\alpha = \tau/v$.
The $n=0$ eigenstate, 
with 
$\epsilon_0 = (y/x_0) \cos \alpha = g \tau$, 
is a zero mode when $\tau = 0$.
Because of this, 
the kink state with $\tau = 0$ is doubly degenerate,
with charge $\half$ or $-\half$,
depending on whether the zero mode is occupied or not \cite{JackiwRebbi76}.
When $\tau \ne 0$, the degeneracy between these states is lifted.

How do fermions change the energy (\ref{kinkenergyxyz}) of the kink,
and what effect does this have on the kink's stability (or lack thereof)?
To answer this, we first compute 
the fermion vacuum energy of the kink (relative to the vacuum),
\begin{equation}
E_{\rm vacuum} (\tau) =  
  -\half \sum_{\epsilon_\lambda > 0} 
   \left[ \epsilon_\lambda ({\rm kink}) - \epsilon_\lambda ({\rm vac}) \right]
  +\half \sum_{\epsilon_\lambda < 0} 
   \left[ \epsilon_\lambda ({\rm kink}) - \epsilon_\lambda ({\rm vac}) \right] 
\end{equation}
summing over all discrete and continuous eigenvalues,
both positive and negative.
(Because of the mode $\epsilon_0$, 
the spectrum is not generally charge-conjugation symmetric.)
We obtain
\begin{equation}
E_{\rm vacuum} (\tau) =  
    -\half g |\tau| + E_{\rm nzm} (\tau)
\end{equation}
where $E_{\rm nzm} (\tau)$ is the contribution from all the non-zero modes
({\it i.e.}, other than $\epsilon_0$).
The appearance of a cusp at $\tau = 0$ is somewhat misleading,
because the kink vacuum refers to the state with the zero mode occupied
for $\tau < 0$ and to the state with the zero mode unoccupied for $\tau> 0$.
The state with the zero mode occupied for $\tau > 0$ 
is an excited state of kink plus particle, and 
the state with the zero mode unoccupied for $\tau < 0$
is an excited state of kink plus hole.
Using eq.~(\ref{effectiveenergyxyz}), we obtain 
\begin{eqnarray}
E_{\rm occupied}(\tau) & = & 
     E_{\rm boson}(\tau) + \half g \tau + E_{\rm nzm} (\tau)\nonumber\\
E_{\rm unoccupied}(\tau) & = & 
    E_{\rm boson}(\tau) - \half g \tau + E_{\rm nzm} (\tau)
\label{occupiedxyz}
\end{eqnarray}
for the energies of the states with the zero mode occupied or unoccupied.
It is possible to compute $ E_{\rm nzm} (\tau) $ in closed form 
by summing the discrete states and integrating
the phase shifts of the continuum modes, 
as in refs.~[\citenum{ChangYan75,Naculich92}].
For example, for $y=1$, one obtains
\begin{equation}
E_{\rm nzm} (\tau)
= {1 \over x_0} \left[
\left( {1\over 2} - {\alpha \over \pi}\right) \cos\alpha 
+ {1\over \pi} \sin\alpha \right].
\end{equation}
(An alternative approach using inverse scattering methods is found in
ref.~\citenum{CampbellLiao76}.)

The actual form of $E_{\rm nzm}(\tau)$ is not important for our purposes;
since all non-zero mode energies 
have vanishing derivative with respect to $\tau$ at $\tau = 0$,
$E_{\rm nzm}(\tau)$ is flat there (as is $E_{\rm boson}(\tau)$).
Hence, because of  the presence of the linear $ \pm \half g \tau $ term
in eqs.~(\ref{occupiedxyz}),
neither of the degenerate kink states 
is an extremum of the energy functional at $\tau = 0$.
The effect of fermions is to shift the saddle point
solution away from the crown of the Mexican hat.

Strictly speaking, 
the kink configurations (\ref{kinkconfigurationsxyz}) are not connected
because their electric charges differ,
being given by 
$Q = -\alpha/\pi = -(1/\pi) \cos^{-1} (\tau/v)$
\cite{GoldstoneWilczek81,JackiwSemenoff83}.
But since the charge of the configuration (modulo integers) 
is solely determined by the field at its endpoints $x = \pm \infty$,
one could analyze instead a family of connected configurations
anchored at $\sigma(\pm \infty) = \pm v$ and $\tau(\pm \infty) =0$
(and therefore all having charge $-\half$),
which progressively slide off the crown of the hat;
our conclusions would remain unchanged.

The situation is somewhat different for the electroweak string.
In that case, 
there is a continuous spectrum of low-lying states
built upon the zero mode (massless fermions running
up and down the string).
When the electroweak string is perturbed, 
all these states contribute to the fermion vacuum energy.
Unlike the kink,
the electroweak string remains an 
extremum of the energy functional,
but it is no longer a minimum of the energy 
for {\it any} values of the parameters
\cite{Naculich95}.
The fermion vacuum energy reduces the stability of the electroweak string.

\section{References}

\vspace{-0.25cm}

\def\PLB{ \it  Phys. Lett.         \bf B}
\def\NPB{ \it  Nucl. Phys.         \bf B}
\def\PRD{ \it Phys. Rev.          \bf D}
\def\PRL{ \it Phys. Rev. Lett.    \bf  }


\begin{thebibliography}{9}

\bibitem{BaggerNaculich91} J. A. Bagger and S. G. Naculich, \PRL 67 \rm (1991) 2252.
\bibitem{BaggerNaculich92} J. A. Bagger and S. G. Naculich, \PRD 45 \rm (1992) 1395.
\bibitem{Naculich95} S. G. Naculich, \PRL  75 \rm (1995) 998.
\bibitem{KonoNaculich96} S. Kono and S. G. Naculich, 
in {\it Particles, Strings, and Cosmology},
J. Bagger, G. Domokos, A. Falk, S. Kovesi-Domokos, eds. 
(World Scientific, 1996, Singapore), 297.
\bibitem{LiuVachaspati96} H. Liu and T. Vachaspati, \NPB 470 \rm (1996) 176.
\bibitem{JackiwRebbi76} R. Jackiw and C. Rebbi, \PRD  13 \rm (1976) 3398.
\bibitem{ChangYan75} S.-J. Chang and T.-M. Yan, \PRD 12 \rm (1975) 3225.
\bibitem{Naculich92} S. G. Naculich, \PRD 46 \rm (1992) 5487.
\bibitem{CampbellLiao76} D. Campbell and Y.-T. Liao, \PRD 14 \rm (1976) 2093.
\bibitem{GoldstoneWilczek81} J. Goldstone and F. Wilczek, \PRL 47 \rm (1981) 986.
\bibitem{JackiwSemenoff83} R. Jackiw and G. Semenoff, \PRL 50 \rm (1983) 439.

\end{thebibliography}
\end{document}